\documentclass[twocolumn,prx,twocolumn,superscriptaddress]{revtex4-1}
\usepackage{amsmath,amssymb,mathrsfs}
\usepackage{natbib}
\usepackage{subfigure}
\usepackage{tabularx}
\usepackage{epsfig}
\usepackage{longtable}
\usepackage{amsfonts}
\usepackage{rotating}
\usepackage{subfigure}
\usepackage{amsmath}
\usepackage{comment}
\usepackage{bbold}

\def\be{\begin{equation}}
\def\ee{\end{equation}}
\def\bea{\begin{eqnarray}}
\def\eea{\end{eqnarray}}
\def\nn{\nonumber}

\def\tbf{\textbf}
\def\up{\uparrow} 
\def\down{\downarrow}
\newcommand*{\Scale}[2][4]{\scalebox{#1}{$#2$}}%
\def\scalefactor{.8}

\usepackage[unicode=true,bookmarks=true,bookmarksnumbered=false,bookmarksopen=false,breaklinks=false,pdfborder={0 0 1},backref=false,colorlinks=true]{hyperref}

\hypersetup{linkcolor=magenta,urlcolor=blue,citecolor=blue,pdfstartview={FitH},hyperfootnotes=false,unicode=true}

\begin{document}
\title{Cross Dimensionality and Emergent Nodal Superconductivity with  $p$-orbital Atomic Fermions}

\author{Xiaopeng Li} \email{xiaopeng\_li@fudan.edu.cn} 
\affiliation{State Key Laboratory of Surface Physics, Institute of Nano-electronics and Quantum Computing, and Department of Physics, Fudan University, Shanghai 200433, China}
\affiliation{Collaborative Innovation Center of Advanced Microstructures, Nanjing 210093, China} 




\begin{abstract} 
I study cross dimensionality of $p$-orbital atomic fermions loaded in an optical square lattice with repulsive interactions. The cross-dimensionality 
emerges when the transverse tunneling of $p$-orbital fermions is negligible. 
With renormalization group analysis, the system is found to support two dimensional 
charge, orbital, and spin density wave states with incommensurate wavevectors. The 
transition temperatures of these states are controlled by perturbations near a one dimensional Luttinger liquid fixed point. 
Considering transverse tunneling, the cross-dimensionality breaks down and the density wave (DW) orders become unstable, and I find 
an unconventional superconducting state mediated by fluctuation effects. The superconducting gap 
has an emergent nodal structure determined by the Fermi momentum, which is tunable by controlling atomic density. Taking an effective description of the superconducting state,  it is shown that the nodal structure of Cooper pairing can be extracted from momentum-space radio-frequency spectroscopy in atomic experiments. These results imply that $p$-orbital fermions could enrich the possibilities of studying correlated physics in optical lattice quantum emulators beyond the single-band Fermi Hubbard model. 
\end{abstract}
\maketitle

{\it Introduction.---} 
Investigation of strongly correlated physics with ultracold atoms has attracted considerable efforts in the last decade. Bose- and Fermi-Hubbard models have been achieved in experiments by loading ultracold atoms in the lowest band of optical lattices~\cite{1998_Zoller_Jaksch_PRL,2002_Hofstetter_Cirac_PRL,2007_Lewenstein_AP,2008_Bloch_Dalibard_RMP,2010_Esslinger_CMP,2015_Lewenstein_RPP}. For bosons, the Mott-superfluid quantum phase transition  has been observed~\cite{2002_Greiner_Mandel_Nature}. For fermions although cooling to the ground state is more experimentally challenging~\cite{2008_Jordens_Strohmaier_Nature,2008_Schneider_Hackermuller_Science},  recent developments~\cite{2013_Greif_Uehlinger_Science,2015_Hart_Duarte_Nature} have mounted to a new milestone with the long-sought anti-ferromagnetic ordered Mott insulator~\cite{2016_Mazurenko_Chiu_arXiv} finally accomplished through the technique of fermionic quantum microscope~\cite{2015_Haller_Hudson_NatPhys,2015_Cheuk_Nichols_PRL,2015_Parsons_Huber_PRL,2015_Edge_Anderson_PRA,2015_Omran_Boll_PRL,2016_Greif_Parsons_Science,2016_Cheuk_Nichols_PRL,2016_Parsons_Mazurenko_Science,2016_Boll_Hilker_Science,2016_Cheuk_Nichols_Science,2016_Brown_Mitra_arXiv}. Upon doping the system is expected to show $d$-wave superconductivity as in cuperates. This opens up a new window to explore correlated physics of fermionic atoms previously thought unpractical. 

Apart from the single-band Fermi Hubbard model, orbital degrees of freedom also play important roles in solid state materials~\cite{2000_Tokura_Nagaosa_Science}. Unconventional superconductivity in iron-based superconductors~\cite{2006_Kamihara_Hiramatsu_JACS}, 
competing orders  in complex oxides and heterostructures~\cite{2009_Takagai_NatMat,2014_Sulpizio_Ilani_ARMR}, 
and the chiral $p$-wave topological state proposed in strontium ruthenates~\cite{1998_Luke_Fudamoto_Nature} 
all have multi-orbital origins.  In optical lattices, ultracold atoms have successfully been put in higher orbitals in search of exotic quantum phases~\cite{2016_Li_Liu_RPP}. For example, a bosonic analogue of the chiral $p$-wave state~\cite{2011_Wirth_Olschlager_NatPhys,2011_Lewenstein_Liu_NatPhys,2015_Kock_Olschlager_PRL}
and an anomalous phase-twisting condensate have been observed~\cite{2012_Soltan-Panahi_NatPhys}. 
In theory, various aspects of $p$-orbital fermions, such as orbital orders~\cite{2006_Wu_Liu_PRL,2008_Zhao_Liu_PRL,2008_Wu_PRL2,2011_Cai_Wang_PRA,2012_Zhang_Li_PRA}, magnetism~\cite{2008_Wu_Zhai_PRB,2010_Zhang_Hung_PRA2,2013_Li_Lieb_PRL}, and topological phases~\cite{2008_Wu_PRL,2008_Wu_Sarma_PRB,2010_Liu_Liu_PRA,2011_Zhang_Hung_PRA,2011_Sun_Liu_NatPhys,2012_Wu_He_PRB,2013_Li_Zhao_NatComm,2014_Liu_Li_NatComm,2014_Dutta_PRA,2016_akrzewski_PRA,2016_Liu_Li_PRA} have been largely investigated. It has been found that spontaneous translational symmetry breaking with incommensurate wavevectors is generally favorable in the system of $p$-orbital fermions. However how the incommensurate DWs  interplay with superconductivity in repulsive $p$-orbital fermions, as analogous to the emergence of $d$-wave superconductivity in the doped anti-ferromagnetic $s$-band Mott insulator, remains  to be understood.


In this work, I study two component (referring to atomic hyperfine states) fermionic atoms, e.g., $^{40}$K loaded in $p$-orbital bands of an optical lattice. 
The orbital cross dimensionality of this system leads to incommensurate spin, orbital and charge density wave phases. Finite transverse tunneling of $p$-orbital fermions will destroy the cross dimensionality and weaken the DW orders, and a nodal superconducting (NSC) state is found to emerge  
when the incommensurate spin density wave (ISDW) is suppressed, which is analogous to the emergence of $d$-wave superconductivity (SC) in the doped 
$s$-band  Mott insulator~\cite{2000_Metzner_PRB,2000_Maier_PRL,2001_Honerkamp_PRB,2011_Raghu_SC_PRB}. 
A crucial difference worth emphasis is that the NSC state in the $p$-orbital setting appears at generic filling instead of 
restricted to near half filling. The nodal structure of the SC pairing is related to Fermi momentum and is  thus tunable via controlling the atom number density.  The spectroscopic properties of the NSC state are predicted, and can be tested with momentum-resolved radio-frequency spectroscopy in atomic experiments.   I also point out that the theory in this work may also shed light on the superconductivity in certain oxide heterostrucutres, e.g.,LaAlO$_3$-SrTiO$_3$, where $d_{xz}$ and $d_{yz}$ orbitals on the interface can be modeled as $p$-orbitals effectively.

{\it Model.---} 
A minimal tight binding model on a square lattice 
describing $p_x$ and $p_y$ orbital fermions is considered here with the 
Hamiltonian $H = H_0 + H_{\rm int}$, where the tunneling term is 
$
H_0 = \sum_{\tbf{r}, \mu, \nu = x,y} t_{\mu \nu} [c_{ \mu \alpha, {\bf r} } ^\dag c_{ \mu \alpha, {\bf r} + \hat{e}_\nu} 
+ h.c.]. 
$
Here the annihilation operators $c_{x, \alpha}$ and $c_{y, \alpha}$  describe the $p_{x}$ and $p_{y}$ orbitals with spin 
polarization $\alpha = \up, \down$, $\hat{e}_{x}$ and $\hat{e}_y$ are two reciprocal lattice vectors for the square lattice, and 
the tunneling matrix is $t_{\mu \nu} = t_\parallel \delta_{\mu \nu} - t_\perp (1-\delta_{\mu \nu} ) $~\cite{2006_Liu_PRA}. 
The lattice spacing is set to be $1$ throughout. 
Assuming spin SU(2) symmetry, the local interaction takes the form 
\bea 
\label{eq:localint}
&& H_{\rm int} = U \sum_{\tbf{r}, \nu} n_{\nu \up} ( \tbf{r} ) n_{\nu \down}  ({\bf r} ) 
		  + U' \sum_{{\bf r}} n_{x} ({ \bf r} ) n_{y} ({\bf r})   \\
&& + J \sum_{{\bf r}} \tbf{S}_x (\tbf{r}) \cdot {\bf S} _y ({\bf r})  
		  + J' \sum_{\bf r} \left[ c_{x \up, {\bf r} } ^\dag c_{x \down, {\bf r} } ^\dag 
		  					c_{y \down, {\bf r} } c_{y \up, {\bf r} } + h.c. \right], \nn 
\eea 
where 
$n_{\nu \alpha} ({\bf r}) = c_{\nu \alpha, {\bf r} } ^\dag c_{\nu \alpha, {\bf r} }$, 
$n_{\nu} ({\bf r}) = \sum_\alpha n_{\nu \alpha} ({\bf r}) $, 
and ${\bf S}_\nu ({\bf r}) = \sum_{\alpha \beta} c_{\nu \alpha, {\bf r} } ^\dag {\bf \sigma} _{\alpha \beta} c_{\nu \beta, {\bf r} }  $ .  
The local interaction 
respects conservation of the fermion parity of each orbital, 
$ \prod_{\bf r} (-1) ^{n_x ({\bf r})}$ and $ \prod_{\bf r} (-)^ {n_y ({\bf r})} $. 
This work focuses on the repulsive interaction with $U>0$. 

The low temperature phase diagram is analyzed  with renormalization group (RG) techniques. 
Due to the fermion parity  and  SU(2) symmetries~\cite{2001_Honerkamp_PRB}, the renormalized interaction takes 
a general form,  
\bea 
\label{eq:renormalint} 
&& \textstyle H_{\rm int} = \frac{1}{2}  \int \prod_{i= 1,2,3} \frac{d{\bf k} _i }{(2\pi)^2 } 	 
		\Scale[\scalefactor]{ 
		V\left(   
		\begin{array}{ccc}
		 {\bf k}_1 & {\bf k}_2 & {\bf k}_3  \\
		 \nu_1 & \nu_2 & \nu_3 
		\end{array} 
	   \right)  
	   } \\ 
&& \textstyle C_{\nu_1 \alpha} ^\dag ({\bf k}_1) C_{\nu_2 \beta} ^\dag ({\bf k}_2) C_{\nu_3 \beta} ({\bf k}_3) 
	   			C_{\kappa(\nu_1, \nu_2, \nu_3) , \alpha}  ({\bf k}_1 + {\bf k}_2 - {\bf k}_3) , \nn
\eea 
where 
$C_{\nu \alpha} ({\bf k})$ is the Fourier transform of $ c_{\nu \alpha, {\bf r}}$, 
and $\kappa$ is a permutation symmetric function defined by 
$\kappa(\nu, \nu, \nu) = \nu$ and $\kappa(\nu, -{\nu}, \nu ) =  - {\nu}$ 
(the `$-$' sign in front of the orbital index $\nu$ means switching between $p_x$ and $p_y$). 
The frequency dependence in the effective interactions is neglected for such dependence is expected to  be 
unimportant~\cite{1994_Shankar_RMP,2001_Honerkamp_PRB,2000_Metzner_PRB}. 
The renormalization of effective interactions is given in Supplementary Information.

\begin{figure}[htp] 
\includegraphics[angle=0,width=.8\linewidth]{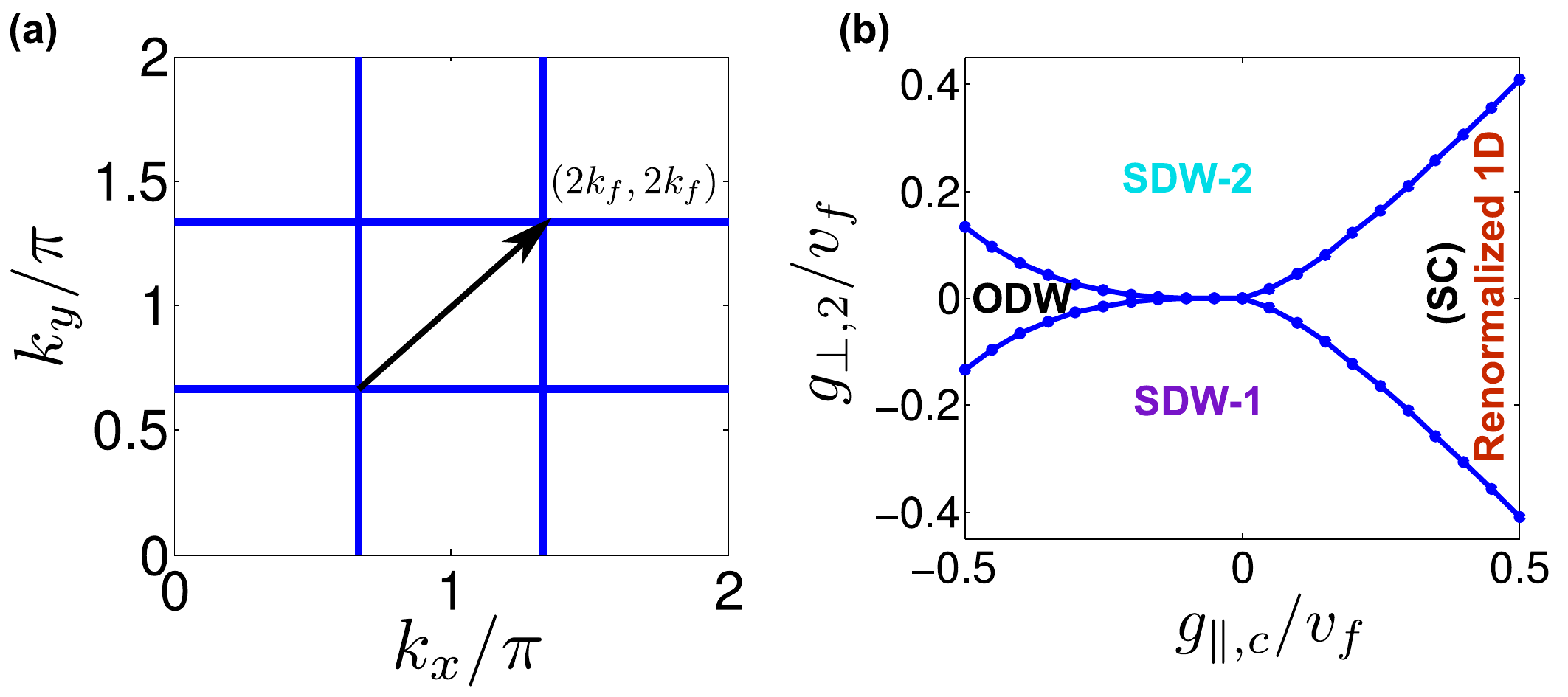}
 \caption{Phase diagram in the strict cross-dimensional limit. 
(a) The illustration of Fermi surfaces at the cross-dimensional limit. (b) The phase diagram 
parameterized by $g_{\perp,c}$ and $g_{\parallel, c}$. 
In SDW-1 (SDW-2), spin polarizations of two orbital components are antiparallel (parallel). 
In this plot, $\tilde{g}_{\perp, 1} = v_f$, 
and ${g_{\parallel, 2}} (q) = v_f$. 
For  $\tilde{g}_{\perp,1}$ with an opposite sign, the ODW state will be replaced by a CDW state. 
}
\label{fig:phasediag} 
\end{figure}

{\it Cross-dimensional limit.---} 
First consider the limit $t_\perp \to 0$. 
In the renormalization to the low-energy limit, the coupling $J'$ is strongly suppressed due to 
the particle-particle channel, provided that $|J'| <U$, which holds for cold atoms with contact interactions. 
I will thus take $J'\to 0$. 
In this limit,  
the system is dynamically cross-dimensional, meaning that 
each orbital tunnels only in the direction along its  own elongation 
whereas they are coupled via inter-orbital interactions. 
Mathematically, the cross-dimensionality is characterized by a  transverse-sliding-phase (TSP) symmetry,  
$c_{x, {\bf r}} \to e^{i \theta(r_y) } c_{x, {\bf r} }$, $c_{y, {\bf r}}   \to  e^{i \theta(r_x) } c_{y, {\bf r}}  $ .
In the absence of inter-orbital interactions, this cross-dimensional system reduces   to 
an ensemble of decoupled one-dimensional systems, each being described by a Luttinger liquid theory. 

Resulting from the one-dimensionality of Fermi surfaces, 
perfect nesting wavevectors ${\bf Q} = (\pm Q, \pm Q)$, with $Q = 2 k_f$ 
(Fig.~\ref{fig:phasediag}(a)). 
This  gives  rise to leading instabilities in DW  channels
\be 
\Scale[\scalefactor]{ 
\textstyle {V} _1 ({\bf k}, {\bf k}') \equiv \textstyle {V} \left( 
			\begin{array}{ccc} 
			{\bf k} & {\bf k}' + {\bf Q} & {\bf k}' \\ 
			\nu 	&-{\nu} 	&-{\nu} 
			\end{array} 
			\right), 
\textstyle {V}_2  ({\bf k}, {\bf k}') \equiv \textstyle {V} \left( 
			\begin{array}{ccc} 
			{\bf k} & {\bf k}' +{\bf Q}   & {\bf k} +{\bf Q} \\ 
			\nu 	&-{\nu} 	&{\nu} 
			\end{array} 
			\right) 
			} .
\nn
\ee  
Divergence of ${V}_2$ in the RG flow leads to formation of 
SDW  orders at low temperature, 
while divergence of $ {V}_2 -2 {V}_1$ leads to charge density 
wave (CDW) or orbital density wave (ODW) orders, depending on the sign of this coupling. 

The low temperature phases are determined by the effective interactions near the Fermi surface. 
The momentum is rewritten in terms of components parallel (${\bf e}_\parallel$) and perpendicular 
(${\bf e} _\perp$) to the Fermi surface as 
 ${\bf k} = k {\bf e}_\parallel + \chi (k_f + l) {\bf e}_\perp$,  
with the chirality $\chi = \pm$ index the 
two Fermi surfaces for each orbital. 
Note that 
${\bf e}_{\parallel}$ and ${\bf e}_{\perp}$ should be defined in an opposite way for 
two orbitals---${\bf e}_{\parallel} = \hat{e}_y$, ${\bf e}_{\perp} = \hat{e}_x$ for $p_x$ and 
${\bf e}_{\parallel} = \hat{e}_x$, ${\bf e}_{\perp} = \hat{e}_y$ for $p_y$. 
Each single-particle mode is now labeled by three indices---parallel momentum $k$, chirality $\chi$, and orbital 
$\nu$. 
The dispersion near the Fermi surface reads 
$
\epsilon_{\nu} ({\bf k}) = v_f l + {\cal O} (l^2),
$ 
with $v_f$ the Fermi velocity. 

The interactions ${V}_1({\bf k, k^\prime})$ and ${V}_2({\bf k, k^\prime})$ projected onto the Fermi surface 
take a more explicit form  
\bea  
\textstyle g_{\perp, 1}  (\nu; \chi, k , \chi',  k' )= 
\textstyle
\Scale[\scalefactor]{ 
 \Gamma \left(  
		\begin{array}{cccc} 
		\textstyle k	&k' -\chi Q		& k' 	&k-\chi' Q  \\ 
		\textstyle \chi	& - \chi' 	&  { \chi'}  &-{\chi}  \\ 
		\textstyle \nu	&-{\nu} 	& -{\nu} 	&\nu 	
		\end{array} 
		 \right) } , \nn 
\eea 
\bea 
 \textstyle g_{\perp, 2}  (\nu; \chi, k, \chi', k') = 
\textstyle 
\Scale[\scalefactor]{ 
\Gamma \left( 
		\begin{array}{cccc} 
		\textstyle k 	& k' - \chi Q 	& k-\chi'Q	& k' \\ 
		\textstyle \chi	&-\chi'	& -{\chi} 	& {\chi'} \\
		\textstyle \nu	&-{\nu} 	&\nu	&-{\nu} 
		\end{array} 
		\right) } ,  \nn 
\eea 
where $\Gamma ( {\rm m}_1, {\rm m} _2, {\rm m}_3, {\rm m}_4)$, with 
${\rm m}_j$ a collective column of indices 
$ \left( 
k_j, 
\chi_j, 
\nu_j 
\right)^T 
$, 
describes scattering from modes ${\rm m}_3$ and ${\rm m}_4$ to ${\rm m}_1$ and ${\rm m}_2$. 
Pictorial illustration of these couplings is given in Supplementary Information. 
Due to the two dimensional nature of these inter-orbital couplings, 
the parallel momentum is not conserved. 
The TSP  symmetry implies    
$
g_{ \perp, j=1,2} (\nu; \chi, k, \chi', k') = g_{ \perp, j} (\nu; \chi, k + p, \chi', k' + p'). 
$ 
The inter-orbital couplings $g_{ \perp, j}$ thus have no $k$ (or $k'$) dependence. Further considering point group $D_4$ symmetry, $g _{\perp, j}$  
does not depend on $\nu$, $\chi$ or $\chi'$.

Under renormalization, the inter-orbital couplings $g_{\perp, j}$ 
are intertwined with intra-orbital ones 
\bea 
\textstyle g_{\parallel, 1} ( \nu, \chi, k, k', q) &=& 
\Scale[\scalefactor]{ \Gamma \left( 
	\begin{array}{cccc} 
	k	& k'+q   & k' & k+q \\ 
	\chi    &-{\chi}	& \chi &-{\chi}  \\ 
	\nu	& \nu 	&\nu	&\nu 
	\end{array} 
\right) }, \nn \\ 
\textstyle g_{\parallel, 2} (\nu, \chi, k, k' , q) &=& 
\Scale[\scalefactor]{ 
\Gamma \left( 
	\begin{array}{cccc}
	k	&k'+q 	&k+q	&k' \\ 
	{\chi}	& -\chi 	& -\chi &{\chi} \\
	\nu	&\nu	& \nu	 &\nu  
	\end{array} 
	\right) } .  \nn 
\eea 
Here, the parallel momentum is conserved, which is different from inter-orbital couplings. 
The other difference comes from the consequence of the TSP symmetry. For the intra-orbital 
couplings, TSP symmetry implies  
$
g_{\parallel, j} (\nu, \chi, k, k', q) = g_{\parallel, j} (\nu, \chi, k+p, k'+p, q). 
$ 
These couplings could therefore have more complicated momentum dependence. 
On the other hand, starting with on-site interactions, the momentum dependence in 
$g_{\parallel, j} $ generated in the renormalization is mainly from intertwined  scattering with inter-orbital couplings, which causes  
the leading momentum dependence of $g_{\parallel, j} (\nu, \chi, k, k', q)$  on the exchange  term $q$, i.e., 
$
g_ {\parallel, j} (\nu, \chi, k, k', q) \approx g_{\parallel, j} (q). 
$  
This is justified by a complete treatment of momentum dependence. 
To explicitly extract the CDW or ODW channel, I introduce  
$
   \tilde{g}_{\perp, 1} = 2 g_{\perp, 1} - g_{\perp, 2},  
   \tilde{g}_{\parallel, 1} (Q) = 2 g_{\parallel, 1} (Q) - g_{\parallel, 2} (Q).  
$

At tree level, the g-ology couplings are related to the lattice model as  
$g_{\perp, 1} = U'-J$, $g_{\perp, 2 } = -2J$, $g_{\parallel, j} = U$. 
From the renormalization at one-loop level (see Supplementary Information), 
the renormalization group (RG) equations of the g-ology couplings  are obtained to be 
\bea 
&& \textstyle  \frac{d \tilde{g}_{\perp,1} } {ds} = \frac{1}{\pi v_f} 
  \left\{ - \tilde{g}_{\parallel, 1} (Q)  \tilde{g}_{\perp, 1}\right\}  \nn \\
&& \textstyle \frac{d g_{\perp, 2} }{ds} = \frac{1}{\pi v_f} \left\{ g_{\perp,2} g_{\parallel, 2} (Q) \right\} \nn  \\
&&\textstyle \frac{d \tilde{g}  _{\parallel, 1} (Q) } {ds }  
 = \frac{1}{2 \pi v_f} \left\{  -\tilde{g} _{\perp,1} ^{ 2} 
-\tilde{g}_{\parallel, 1}^2 (Q) 
 \right.   \\ 
&&\textstyle  \left. \,\,\,\,\,\,\,\,\,  		
+ \overline{g_{\parallel,1} ^2 (q) } + \overline{g_{\parallel, 2} ^2 (q)  } 
-4 \overline{g_{\parallel, 1} (q)  g_{\parallel,2}(q) }  				
\right\} \nn \\
&& \textstyle \frac{d g_{\parallel,2} (Q)}{ds} = \frac{1}{2\pi v_f} \left\{ g_{\perp,2} ^{2}  + g_{\parallel,2} ^{ 2}  (Q) 
 - \overline{ g_{\parallel,2} ^{2} ( {q} )}  - \overline{g_{\parallel,1} ^{  2} ({q})}   \right\} . \nn 
\eea 
A shorthand notation is adopted $\overline{f(q) } = \int \frac{d q}{2\pi} f(q)$.  
The RG low of $g_{\parallel, j} (q)$ for the momentum $q$ away from $Q$ is obtained as   
\bea 
 &&  \textstyle \frac{d  g_{\parallel, 1}  (q )  } {ds }  
= \frac{1}{\pi v_f}  \left\{ -g_{\parallel,1} ^ { 2 } (q)  + g_{\parallel,1} (q) g_{\parallel,2} (q)  \right.\nn \\	
&& \,\,\,\,\,\,\,\,  \,\,\,\,\,\,\,\, \left. \textstyle - \overline{ g_{\parallel, 1} (p) g_{\parallel, 2}  (p)  } \right\},  \nn \\ 
&& \textstyle \frac{d g_{\parallel, 2} (q)  }{ds} 
 =\frac{1}{2\pi v_f } \left\{   g_{\parallel,2} ^{2}  (q) 
-\overline{ g_{\parallel,2} ^{2} (p)  } - \overline{  g_{\parallel,1}^{  2} (p) } \right\}. 
\eea 
The RG flow of  intra-orbital couplings with $q \neq Q$ is  not affected by inter-orbital couplings. 
(More rigorously, $q\neq Q$ means $|q-Q| > \Lambda$, with $\Lambda \ll 2\pi$ for weak interactions.)

Since the strong momentum dependence in $g_{\parallel, j}$ near $q \approx Q$ is restricted to very limited phase space, I  
approximate $\overline { g_{\parallel, j} (q)  g_{\parallel, j'} (q) }$ by 
$[\overline { g_{\parallel, j} (q) }] [\overline{  g_{\parallel, j'} (q) }] $, and the deviation is quantified by 
$\varepsilon_{jj'} = \overline{\Delta g_{\parallel, j} (q) \Delta g_{\parallel, j'} (q) } $, with 
$
\Delta g_{\parallel, j} (q) = g_{\parallel, j} (q) -  \overline{g_{\parallel, j} }.  
$
The solution for $\overline{g_{\parallel, j} }$ is obtained to be 
$ 
 [\overline{g_{\parallel, 1}}  (s ) ]^{-1}  = [ \overline{g _{\parallel, 1}} (0)]^{-1}  + \frac{s}{\pi v_f} + {\cal O} (\varepsilon) 
 $,  
$
 \overline{g_{\parallel, 2}} (s)  =  \overline{g _{\parallel, 2}}  (0) + \frac{1}{2} \overline{g_{\parallel, 1}} (s) 
  - \frac{1}{2} \overline{g_{\parallel, 1} } (0) . 
$  
With repulsive interaction, the intra-orbital coupling $\overline{g_{\parallel, 1}}$ will renormalize to $0$, 
and $\overline{g_{\parallel, 2}}$ will renormalize 
to a constant because 
$
g  _{\parallel,c} \equiv \frac{1}{2} \overline{g_{\parallel, 1}} (0) - \overline{g_{\parallel, 2}} (0)  
$ 
is invariant in the RG flow. 
This leads to a fixed line at $ (g_{\perp, 1} = g_{\perp,2} =0$, ${g_{\parallel, 1}} = 0$, 
$ {g_{\parallel,2}}  = -{g}_{\parallel,c})$, which 
corresponds to a critical Luttinger liquid phase. 
Around this fixed line, the RG flow is determined by 
$ \frac{d}{ds} \tilde{g}_{\perp,1} \propto -  g_{\parallel,c} \tilde{g}_{\perp,1}  $, 
$\frac{d}{ds} g_{\perp, 2}  \propto -g_{\parallel,c}  g_{\perp, 2}$, 
and 
$\frac{d}{ds} \Delta g_{\parallel, j} (q) \propto - g_{\parallel,c} \Delta g_{\parallel, j} (q)  $. 
The two cases with $g_{\parallel, c} <0$ and $g_{\parallel, c} >0$ are different.

On-site repulsion leads a negative  $g_{\parallel, c}$, for which the Luttinger liquid phase is unstable towards DW orders.
Which DW is favorable at low temperature is determined by the comparison among 
these couplings, $\tilde{g}_{\perp,1}$, ${g}_{\perp, 2}$, and $\Delta g_{\parallel, j} (q)$.  
For $p$-orbital fermions, the inter-orbital interactions $\tilde{g}_{\perp,1}$ and $g_{\perp,2}$ are more likely to be dominant, since 
they are related to local interactions. 
A dominant coupling $g_{\perp,2}$ would support $(2k_f, 2k_f)$-SDW orders, in which 
spin polarizations in both $p_x$ and $p_y$ fermions exhibit finite ground-state expectation values, $\langle {\bf S}_x \rangle $ and $\langle {\bf S}_y \rangle$,  and 
oscillate with a wavevector ($2 k_f , 2k_f)$.  
For $g_{\perp,2}<0$  ($g_{\perp,2} >0$), the spin polarizations $\langle {\bf S}_x \rangle $ and $\langle {\bf S}_y \rangle$ 
are  antiparallel (parallel). 
Analyzing the RG flow (see Supplementary Information) the 
scaling of transition temperatures, $T_{{\rm c},1}$ for ODW/CDW and 
$T_{{\rm c},2}$ for SDW is obtained to be,  
$
T_{{\rm c},1  } \propto \left| \tilde{g}_{\perp,1} \right|^{\pi v_f/g_{\parallel,c} }$,   
$ T_{{\rm c},2  } \propto\left| g_{\perp, 2}\right|  ^ {\pi v_f/g_{\parallel,c}}.
$

To complete the theory, I also discuss the possibilities with momentum dependent intra-orbital couplings  
to be dominant. 
If $\Delta g_{\parallel, 2}(q)$ is dominant 
and peaked at some momentum $q_0$, a $(q_0, 2k_f)$-SDW order is favorable  ($\langle {\bf S}_x\rangle $ and 
$\langle {\bf S}_y\rangle $  would oscillate with wavevectors ($2k_f$, $q_0$) 
and ($q_0$, $2k_f$), respectively). For example, with nearest neighbor spin exchange interactions, 
$
\textstyle J_{nn}  \sum_{\bf r} \left[ \vec{S}_x ({\bf r}) \cdot \vec{S}_x ({\bf r}+\hat{e}_y) + x\leftrightarrow y\right],
$ 
$\Delta g_{\parallel, 2}$ (q) could be 
peaked at $q_0 = \pi$ and the ground state then has a $(\pi, 2k_f)$-SDW order, which is consistent with 
previous Bosonization analysis of coupled 1D chains~\cite{1976_Klemm_PRB,1983_Schulz_JPHYSC}. 
Otherwise if $\Delta g_{\parallel, 1} (q_0) $ is dominant, a $(q_0, 2k_f)$-CDW/ODW order is favorable.

With $g_{\parallel,c}  >0$, the Luttinger liquid fixed line is stable against infinitesimal perturbations in DW channels discussed above, and a renormalized one dimensional Luttinger liquid state emerges in the bare two-dimensional system. 
But when the magnitude of the inter-orbital coupling $g_{\perp, 2}$ is larger than some critical value, 
$|g_{\perp,2}  | >g_{\rm} ^{c}$, 
RG flow would escape the attraction  of the stable fixed point. 
Finite $g_{\perp,2} $ thus lead to phase transitions towards two dimensional SDW states (Fig.~\ref{fig:phasediag}(b)). 
One remark is that the renormalized one dimensional state is only stable in the strict cross-dimensional limit. 
Including $J'$ would generate an effective coupling in the SC channel. The system then develops superconductivity at very low temperature.

\begin{figure}[htp] 
\includegraphics[angle=0,width=.8\linewidth]{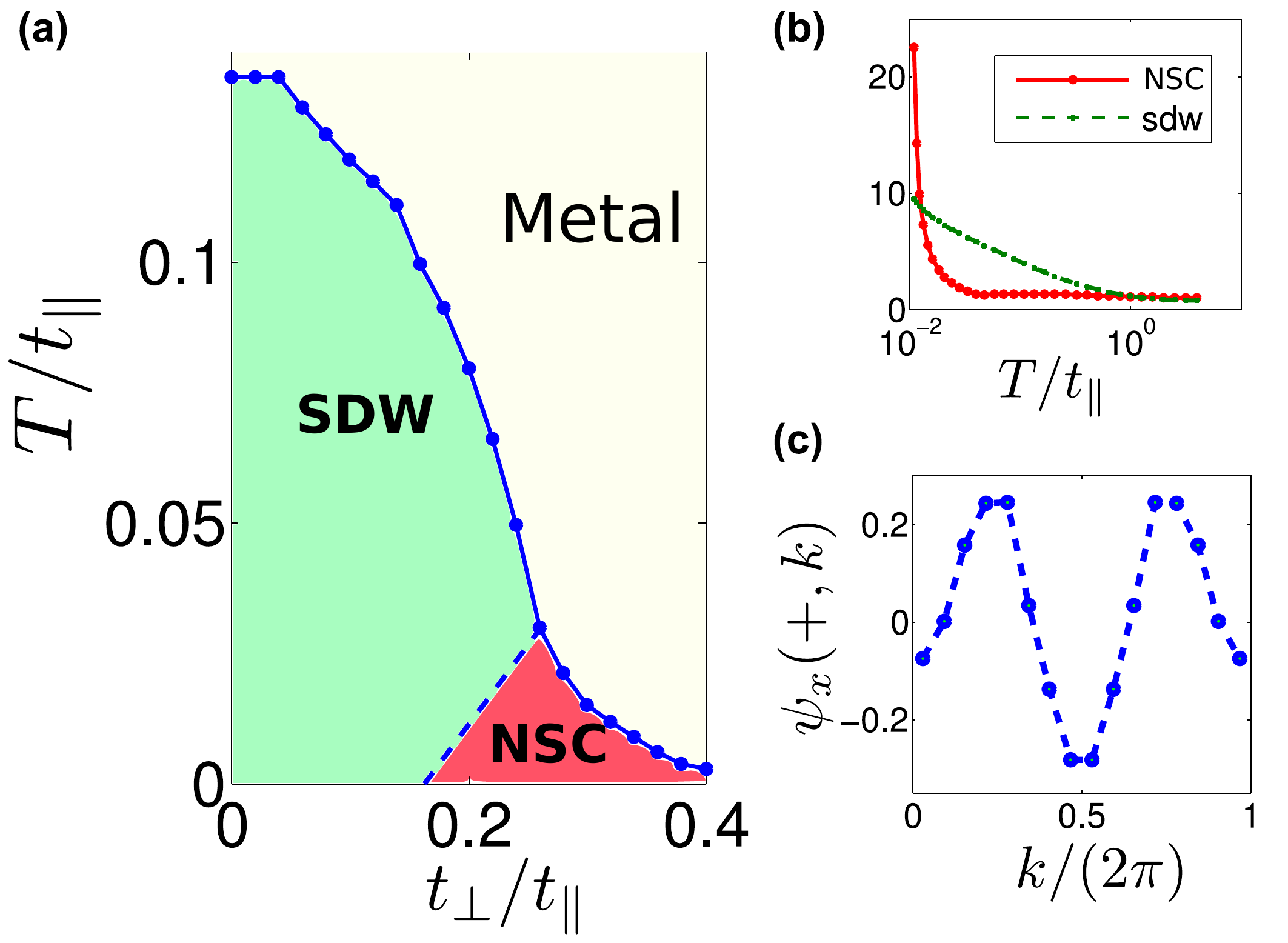}
 \caption{Phase transition from spin density wave (SDW) to a nodal superconducting (NSC) state 
with the model Hamiltonian (Eq.~\eqref{eq:localint}). In this plot I choose 
$U/t_{\parallel} = 2$, $J/t_{\parallel} = 0.5$, 
 and the Fermi level is fixed at $-t_{\parallel}$.  
(a) shows finite temperature phase diagram. The `solid' line is from functional RG (FRG) calculation. 
The SDW order gets weaker with increasing transverse tunneling $t_{\perp}$.
Above a critical value $t_{\perp} ^{\rm c} \approx 0.26 t_{\perp}$, 
the SC instability becomes dominant. 
From the FRG results, it is expected that there is a phase transition from SDW to NSC marked by the `dashed' line.  
  (b) shows the RG flow of the instability strength 
in SC and SDW channels for $t_{\perp}/t_{\parallel} = 0.32$. 
(c) shows the momentum dependence of the pairing on one side of $p_x$ Fermi surface (see main text). 
}
\label{fig:scdome} 
\end{figure}


\paragraph*{Transverse tunneling and a nodal superconducting state.} 
With finite transverse tunneling $t_\perp \neq 0$, the perfect Fermi surface nesting no longer holds and the g-ology RG description breaks down, 
nonetheless the SDW states could be stabilized by finite repulsive interactions. 
A functional RG approach is adopted to approximately solve the full renormalization equation (see Supplementary Information) by a patching scheme~\cite{2001_Honerkamp_PRB}, in which the interaction (Eq.~\eqref{eq:renormalint}) 
is approximated by its projection to the Fermi surfaces. 
To determine the transition temperature more precisely, a temperature renormalization 
scheme is implemented, where the renormalized interactions  explicitly represent temperature dependence of 
effective scatterings of low-energy modes~\cite{2001_Honerkamp_Salmhofer_PRB}.

In the calculation, the SDW and SC channels---$ V_2 ({\bf k}, {\bf k}') $ and 
\be 
\textstyle
\Scale[\scalefactor]{
 V_{\rm SC} (\nu, {\bf k}; \nu', {\bf k}')  = 
V\left(   
		\begin{array}{ccc}
		 {\bf k} & -{\bf k} & -{\bf k}'  \\
		 \nu  & \nu & \nu' 
		\end{array} 
	   \right) } . 
\nn 
\ee 
The strengths of instabilities are characterized by their eigenvalues $\lambda_{\rm SDW}$ and 
$\lambda_{\rm SC}$ of largest magnitude. 
 The low temperature phase 
is determined by which channel is the most divergent. 
The transition temperature is the point where the most dominant channel diverges (in numerics, 
it is determined by which eigenvalue $|\lambda|$ reaches $20 t_{\parallel}$ first.)   
The phase diagram is shown in Fig.~\ref{fig:scdome}. 

The momentum dependence of the SC pairing structure function, 
$
 \langle C_{\nu \uparrow } ( {\bf k}) C_{\nu \downarrow} (-{\bf k})  \rangle, 
$ 
is related to  the eigenvector $\psi_\nu ({\bf k} )$ of $V_{\rm SC}$. 
Rewriting the momentum on the Fermi surface in terms of parallel momentum ($k$) and chirality ($\chi$), the paring function follows  
$
 \psi_\nu (\chi, k) \propto \langle C_{\nu \uparrow } (\chi, k ) C_{\nu \downarrow} (-\chi, -k) \rangle . 
$ 
The relative sign between $\psi_x $ and $\psi_y $ is determined by the sign of $J'$. 
The pairing structure function is found to exhibit nontrivial momentum dependence as shown in 
Fig.~\ref{fig:scdome}(c).  At each Fermi surface, 
there are four nodal points in $\psi_\nu (\chi, k)$, approximately located at 
\be 
k_{\rm node}  = \pm k_f, \pi\pm k_f,
\label{eq:knodes} 
\ee  
with $k_f$ the Fermi momentum defined at the limit of $t_\perp \to 0$.  
The emergence of this nodal structure could be understood by considering the overlap channels between 
$V_{\rm SC}$ and $V_2$, 
$
V_{\rm SC} (\nu, {\bf k}; \nu, {\bf k}) = V_2 ({\bf k}, {\bf k} ), 
$ 
for those special momenta ${\bf k}$ as described in Eq.~\eqref{eq:knodes}. 
In the SC state, the system still has strong tendency towards forming 
SDWs with the wavevector ${\bf Q} \approx (2 k_f, 2k_f)$. 
This means there are large and positive,  but otherwise 
not divergent, couplings in $V_2$. 
Due to the overlap between the SDW and SC channels, these special points form ``hot spots'' for 
SC pairing, which give rise to nodal points  in  
the ground state eigenvector of $V_{\rm SC} (\nu,{\bf k}; \nu' {\bf k}')$. 
In the aspect of spin symmetry, this SC state is a singlet. 

\begin{figure}[htp] 
\includegraphics[angle=0,width=\linewidth]{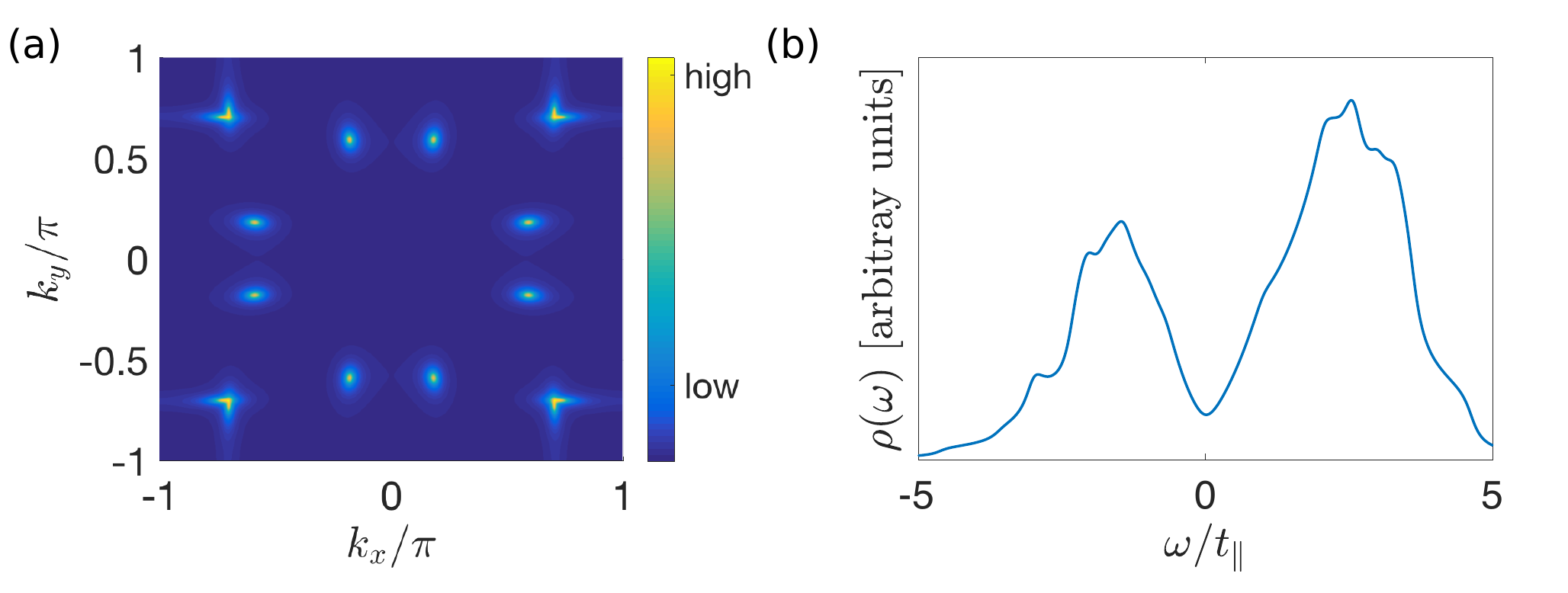}
 \caption{Experimental signatures of the nodal superconductivity of p-orbital fermions. (a) shows the spectra function $A( {\bf k}, \omega)$ at zero energy $\omega = 0$ in arbitrary units, and (b)  the density of states $\rho (\omega) = \sum_{\bf k} A( {\bf k}, \omega)$. In (a), the peaks of $A({\bf k}, 0)$ correspond to the nodal points of the superconducting pairing at the fermi surface in the momentum space. 
 The `V'-shape feature of $\rho (\omega)$ at zero energy shown in (b)  is a signature of gapless excitations.   
   } 
\label{fig:nodalspectra} 
\end{figure}

\paragraph*{Experimental signatures of the nodal superconducting state. }
To observe the NSC state in atomic experiments, spectroscopic properties are studied through an effective description. Deep in the NSC phase, the low energy physics is expected to be described by an effective Bogoliubov de-Gennes (BdG) Hamiltonian, 
\bea 
\textstyle H_{\rm BdG} 
= \sum_{\nu \bf k} && \left\{   \epsilon_\nu  ({\bf k})  C_{\nu \alpha } ^\dag ({\bf k}) C_{\nu \alpha} ({\bf k}) \right.     \nn \\ 
&& \textstyle + \left.   [\Delta ^* ( k_{-\nu} ) C_{\nu \uparrow } ( {\bf k}) C_{\nu \downarrow } (-{\bf k})  + H.c.] \right\}. 
\label{eq:HBdG}
\eea  
Here the dependence of Cooper pairing on the momentum perpendicular to the Fermi surface is neglected for simplicity. Owing to the form of the Cooper pair observed in Fig.~\ref{fig:scdome}, I use an ansatz $\Delta (k) = \Delta_1 \cos (k) + \Delta _2 \cos (2k)$. Keeping higher harmonics is more precise but does not change the physics to be presented below. 

Taking the BdG Hamiltonian, the momentum resolved spectra function $A({\bf k}, \omega)$ is calculated through Green function methods~\cite{2010_Altland_Simons_Book}. The momentum dependence of $A({\bf k}, \omega)$ at zero energy is shown in Fig.~\ref{fig:nodalspectra} (a), 
where the peaks of $A({\bf k}, 0)$ reveal nodes of the SC pairing. The `V'-shape feature of the density of states  $\rho(\omega) = \sum_{\bf k} A({\bf k}, \omega)$ shown in Fig.~\ref{fig:nodalspectra}(b) is a signature of  the gapless Bogoliubov quasi-particles near the nodal points of SC gap. These properties can be tested in momentum-space resolved radio-frequency spectroscopy~\cite{2003_Gupta_Hadzibabic_Science,2003_Regal_Jin_PRL,2007_Sin_Schunck_PRL,2009_Chen_He_RPP}. 


\paragraph*{Conclusion.}
I have derived cross-dimensional g-ology RG flow for $p$-orbital fermions. Charge, orbital, spin density waves and 
their transition temperatures are described  within our g-ology study at the limit of vanishing transverse tunneling.  At 
finite transverse tunneling, DW orders are found to be suppressed, giving rise to an unconventional nodal superconducting 
state.  

\paragraph*{Acknowledgement.}
The author thanks helpful discussions with W. Vincent Liu and Bo Liu. The work is supported by the Start-Up Fund of Fudan University. 

\bibliography{porbfermion}
\bibliographystyle{apsrev4-1}

\newpage

\begin{widetext}

\renewcommand{\thesection}{S-\arabic{section}}
\renewcommand{\theequation}{S\arabic{equation}}
\setcounter{equation}{0}  
\renewcommand{\thefigure}{S\arabic{figure}}
\setcounter{figure}{0}  

\section*{\Large\bf Supplementary Information}

\section{Renormalization equation} 
The renormalization of couplings upon integrating out high-energy modes is given in this section. 
is described by a flow equation~\cite{2001_Honerkamp_PRB,2000_Metzner_PRB}
\bea 
\label{eq:generalRG} 
&& \Lambda \frac{d}{d \Lambda} 
	\Scale[\scalefactor]{ {V} \left(   
		\begin{array}{ccc}
		 {\bf k}_1 & {\bf k}_2 & {\bf k}_3 \\
		 \nu_1 & \nu_2 & \nu_3 
		\end{array} 
	   \right) }   \\
&=& \sum 
	   	\dot{\Pi}_{\rm pp} \left(\epsilon_\mu ({\bf q}) , \epsilon_{\mu_{\rm pp}} ({\bf q}_{\rm pp}) \right) 
	\Scale[\scalefactor]{	{V} \left(   
		\begin{array}{ccc}
		 {\bf k}_1 & {\bf k}_2 & {\bf q} \\
		 \nu_1 & \nu_2 & \mu 
		\end{array} 
	   \right)  
	   	{V} \left( 
		\begin{array}{ccc}
		 {\bf q}_{\rm pp} &{\bf q}  & {\bf k}_3\\
		 \mu_{\rm pp} & \mu & \nu_3   		 
		\end{array} 
		\right) }  \nn \\
& -&  \sum 
		\dot{\Pi} _{\rm ph} \left( \epsilon_{\mu} ({\bf q}), \epsilon_{\mu_{\rm ph} } ({\bf q}_{\rm ph} ) \right) 
		\left\{ 
		 -2 \Scale[\scalefactor]{ {V}  \left( 
		  \begin{array}{ccc}
		 {\bf q}  & {\bf k}_2 & {\bf k}_3 \\
		 \mu & \nu_2 & \nu_3 
		\end{array} 
	   	\right)   
		 {V} \left( 
		  \begin{array}{ccc}
		 {\bf k}_1  & {\bf q}_{\rm ph} & {\bf q}\\
		 \nu_1 & \mu_{\rm ph} & \mu
		\end{array} 
	   	\right)  } \right. \nn \\
&+ & \Scale[\scalefactor]{ \left.		 V \left( 
		  \begin{array}{ccc}
		 {\bf q}  & {\bf k}_2 & {\bf k}_3 \\
		  \mu & \nu_2 & \nu_3 
		\end{array} 
	   	\right)   
           {V} \left( 
		  \begin{array}{ccc}
		 {\bf q}_{\rm ph}  & {\bf k}_1   & {\bf q}\\
		\mu_{\rm ph}   &  \nu_1 & \mu
		\end{array} 
	   	\right)   
		+ {V} \left( 
		  \begin{array}{ccc}
		 {\bf k}_2 & {\bf q}  & {\bf k}_3 \\
		  \nu_2 & \mu & \nu_3 
		\end{array} 
	   	\right)   
		 {V} \left( 
		  \begin{array}{ccc}
		 {\bf k}_1  & {\bf q}_{\rm ph} & {\bf q}\\
		 \nu_1 & \mu_{\rm ph} & \mu
		\end{array} 
	   	\right)  
		\right\} }  \nn \\
&- & \sum 
	\dot{\Pi}_{\rm ph} \left(\epsilon_{\mu} ({\bf q}), \epsilon_{\mu_{\rm ph} ' } ({\bf q}_{\rm ph} ' )  \right) 
\Scale[\scalefactor]{	{V} \left( 
		  \begin{array}{ccc}
		{\bf k}_1 & {\bf q}  & {\bf k}_3 \\
		 \nu_1 & \mu & \nu_3 
		\end{array} 
	   	\right)   
	 {V} \left( 
		  \begin{array}{ccc}
		{\bf q}_{\rm ph} ' & {\bf k}_2  & {\bf q} \\
		\mu_{\rm ph} '  & \nu_2 & \mu 
		\end{array} 
	   	\right)   }.  \nn 
\eea 
Here  $\sum$ means $  \sum_\mu  \int \frac{d^2 {\bf q} }{(2\pi)^2 }$, 
${\bf q}_{\rm pp} = {\bf k}_1 + {\bf k}_2 - {\bf q}$, 
${\bf q}_{\rm ph} = {\bf q} + {\bf k}_2 - {\bf k}_3 $, 
${\bf q}_{\rm ph} ' = {\bf q} +{\bf k}_1 - {\bf k}_3$, 
$\mu_{\rm pp} = \kappa ( \nu_1, \nu_2, \mu)$, 
$\mu_{\rm ph} = \kappa (\nu_2, \nu_3, \mu)$, 
$\mu_{\rm ph}' = \kappa (\nu_1, \nu_3, \mu)$. 
The particle-particle/hole functions are 
$$
\dot{ \Pi}  _{\rm ph (pp) } (\epsilon, \epsilon')  =
\Lambda \frac{\Theta (\mp \epsilon \epsilon') }{|\epsilon| + |\epsilon'|}
\left\{ \Theta ( |\epsilon|  -\Lambda ) \delta (|\epsilon'|-\Lambda) + \epsilon \to \epsilon'  \right\},   
$$ 
with $\Theta(x)$ the heavyside step function.

\section{Transition temperatures of density wave states} 
The transition temperatures of DW states are estimated to be 
\be 
T_c \approx \Lambda_0 e^{-s^*}, 
\ee 
with $s^*$ the point where the corresponding coupling diverges. In this section, 
 the bare inter-orbital interactions are assumed to be much weaker compared to intra-orbital ones. 

To calculate the transition temperature for SDW, the RG flow  
of $g_{\perp, 2}$ is derived assuming the intra-orbital couplings are at the Luttinger liquid fixed point, i.e., 
$g_{\parallel,2}(q) = -g_{\parallel,c}, {g}_{\parallel,1} = 0$.  
(if the bare couplings are not at the fixed point, the bare couplings can simply be replaced by 
renormalized ones). 
Then the RG equation reads, 
\be 
\frac{d}{ds} \left[ g_{\parallel,2} (Q) \pm g_{\perp,2} \right] 
= \frac{1}{2\pi v_f} \left\{ \left[ g_{\parallel,2} (Q) \pm g_{\perp,2} \right] - g_{\parallel, c} ^2 \right\}, 
\ee 
from which it follows 
\be 
g_{\parallel,2} (Q; s) + | g_{\perp,2}(s) | 
= g_{\parallel,c} \frac{\alpha + e^{ s g_{\parallel, c} /(\pi v_f) } }{ \alpha - e^{ s g_{\parallel, c} /(\pi v_f) } }, 
\ee 
with $\alpha$ defined by 
$
\frac{\alpha + 1}{\alpha-1} = \left[ g_{\parallel,2} (Q; 0) + | g_{\perp,2} (0) | \right] /g_{\parallel,c}. 
$ 
For $|g_{\perp,2}| /g_{\parallel,c} \ll 1$, 
$ 
\alpha^{-1} = \frac{1}{2} \frac{g_{\perp,2}}{g_{\parallel,c}} 
+ {\cal O} \left( ( \frac{g_{\perp,2}}{g_{\parallel,c}})^2 \right)  . 
$ 
The scale $s^*$ is then obtained to be 
$
e^{-s^*} \approx \left( \frac{1}{2} \frac{|g_{\perp,2}| }{g_{\parallel,c} }  \right)^ {\pi v_f/g_{\parallel,c} } , 
$ 
from which the scaling of transition temperature for SDW follows, 
$T_{{\rm c},2  } \propto\left| g_{\perp, 2}\right|  ^ {\pi v_f/g_{\parallel,c}}$. 
With similar analysis for RG flow of $\tilde{g}_{\perp,1}$, 
the transition temperature for CDW/ODW is obtained as,  
$T_{{\rm c},1  } \propto\left| \tilde{g}_{\perp, 1}\right|  ^ {\pi v_f/g_{\parallel,c}}$.

\begin{figure}[htp] 
\includegraphics[angle=0,width=\linewidth]{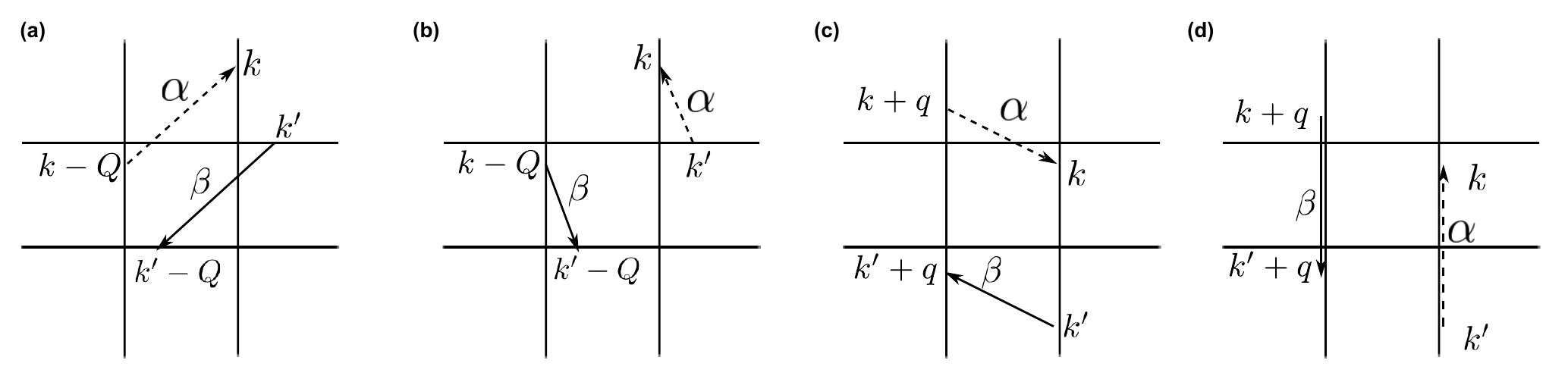}
 \caption{Illustration of g-ology couplings. (a), (b), (c) and (d) show $g_{\perp,1} (p_x, +, k, +, k')$, 
 $g_{\perp,2} (p_x, +, k, +, k')$, $g_{\parallel,1} (p_x, +, k, k', q)$ and $g_{\parallel, 2} (p_x, +, k, k', q)$, 
 respectively (see main text).  } 
\label{fig:gcouplingillustrate} 
\end{figure}

\end{widetext}

\end{document}